\documentclass[pdflatex,sn-mathphys-num]{sn-jnl}

\usepackage{graphicx}%
\usepackage{multirow}%
\usepackage{amsmath,amssymb,amsfonts}%
\usepackage{amsthm}%
\usepackage{mathrsfs}%
\usepackage[title]{appendix}%
\usepackage{xcolor}%
\usepackage{textcomp}%
\usepackage{manyfoot}%
\usepackage{booktabs}%
\usepackage{algorithm}%
\usepackage{algorithmicx}%
\usepackage{algpseudocode}%
\usepackage{listings}%

 \usepackage{bbm}

\newcommand{\be}{\begin{equation}}
\newcommand{\ee}{\end{equation}}

\newcommand{\beqa}{\begin{eqnarray}}
\newcommand{\eeqa}{\end{eqnarray}}
\newcommand{\nn}{\nonumber}

\def\CD {{\cal D}}

\def\CO {{\cal O}}

\def\CR {{\cal R}}

\theoremstyle{thmstyleone}%
\theoremstyle{thmstyletwo}%

\theoremstyle{thmstylethree}%

\begin{document}

\title[Hydrogen Partition Function]{  Renormalisation of  Hydrogen Partition Function \\
and Saha Ionisation Equation


  
}

\author[1]{\fnm{George} \sur{Savvidy}}\email{savvidy@inp.demokritos.gr}

\author[2]{\fnm{Pengming} \sur{Zhang}}\email{zhangpm5@mail.sysu.edu.cn}


\affil[1]{\orgdiv{Institute of Nuclear and Particle Physics}, \orgname{NCSR "Demokritos"}, \orgaddress{\street{Ag. Paraskevi}, \city{Athens}, \postcode{15342},  \country{Greece}}}

\affil[2]{\orgdiv{School of Physics and Astronomy}, \orgname{ Sun Yat-Sen University}, \orgaddress{\street{Daxuelu}, \city{Zhuhai}, \postcode{519082}, \country{China}}}

\abstract{  An important physical quantity that is part of the Saha ionisation equation is the partition function of the neutral hydrogen atom.  An immense uncertainty is caused  by the divergency of the neutral atom partition function arising due to the fast growing degeneracy factor of its excited n-state. 
We suggested a regularisation and renormalisation procedure  for the partition function of the neutral hydrogen atom. The renormalised partition function is well defined and allows to calculate the population of the energy levels and the ionisation state of the hydrogen plasma as a function of its temperature and pressure.  In the low-temperature limit it smoothly transforms to the Saha equation and at  higher temperatures  it describes the  deviation from the  Saha equation.  The application to the solar core and solar atmosphere is considered.  
  }




\maketitle

\newpage

\section{\it Introduction }
 
The Saha ionisation equation  \cite{Saha-1920,Saha-1920-a, Saha-1921,Saha_1921-b}  describes the equilibrium between electrons, ions, and neutral atoms in a plasma and enables to interpret the  physical conditions of stars and nebulae by relating the degree of ionisation to the temperature and pressure  \cite{fowler1923dissociation, Fowler1,Fowler2, 10.1093/mnras/stz549, Blanco2014, Eisenstein_2011,Majewski_2017,Solar_spectra2023} and in cosmology it describes the cosmic reionisation of hydrogen and helium  \cite{1987ApJ...319..195D, 1982ApJ...258..404U, 1985ApJ...290...24E, 1975BAAS....7..366E,1936Phy.....3..729U, 1986ApJ...310..723R, 1957ApJ...125..742E,1965ZNatA..20..184D, 1965dplt, Larkin, 1983ApJ...272..377R, PLASTINO2019122169, Zier:2026mem,KRAMIDA2010586,Savvidy:2026yym}.

The rigorous way of deriving the ionisation formula, which Saha had obtained earlier, has been  derived by Fowler using the theory of molecular dissociation \cite{fowler1923dissociation, Fowler1,Fowler2}.  Attention has been given to the description of electrons according to the Fermi-Dirac statistics by Chandrasekhar and Milne \cite{Chandrasekhar_1931MNRAS, milne1930dissociation}.  

An important physical quantity in the Saha equation introduced by Fowler was the inclusion of the effect of the excited states of neutral hydrogen atoms by the partition function  $ZI(T)$. The   partition function of $H,H^+, e^-$ plasma is the product of the partition functions of the individual contributions of electrons and of the neutral and ionised hydrogen states.   The inclusion of the partition function $ZI(T)$  into the Saha equation introduced   uncertainty that is caused  by the divergency of the $ZI(T)$ arising due to the fast growing degeneracy factor $n^2$ of atom excited n-state  \cite{Planck,Fowler1,Fowler2, Urey1924, Fermi1, Fermi2, Uhlenbeck, PhysRev.94.77} (a short review of divergency problem is given in  Appendix A). The divergency of $ZI(T)$  appears as a real difficulty of the existing theory.

We suggest the epsilon regularisation and the corresponding  renormalisation  of the neutral hydrogen partition function by adopting the methods that were well developed  in the quantum field theory and  strings \cite{Brink:1973oxp}. The renomalised partition function of the neutral hydrogen atom we obtained has the following from:
\be\label{mainpart}
 ZI(T) = z_0 e^{-A}  \sum^{\infty}_{m=2}   ~ \zeta(2m-2) \frac{A^m}{m!}, ~~~~A = \frac{Ry}{kT}, 
\ee
where  $Ry$ is the Rydberg constant, $\zeta(s)$ is the Riemann zeta function and $z_0$ is the internal partition function of the hydrogen atom. In Saha ionisation equation the $Z(T)$ is approximated by the ground state degeneracy  $ZI_{Saha}= z_0$,   assuming that most of the neutral hydrogen is in the ground state.  Therefore we define the ratio 
\be
\CR = \frac{ZI}{ZI_{Saha}} = e^{-A}  \ \sum^{\infty}_{m=2}   ~ \zeta(2m-2) \frac{A^m}{m!} 
\ee
to measure  the deviation from the Saha equation.  The behaviour of $\CR$ as a function of temperature is shown in Fig.\ref{fig1}.  The result is applicable to the deuterium, tritium and hydrogen-type systems.  It would be important to measure the deviation $R(T)$ in the high-temperature region in the laboratories or Tokamak fusion facilities \cite{TokamakChina}. 

\section{\it Renormalisation of Hydrogen Partition Function}

The partition function of neutral hydrogen HI atom is defined as  \cite{Planck,Fowler1,Fowler2, Urey1924, Fermi1, Fermi2, Uhlenbeck, PhysRev.94.77}
\beqa\label{bare}
ZI_{bare} = \sum^{\infty}_{n=1} z_0 n^2 e^{-\frac{E_{n1}}{k T} } , ~~~~E_{n1} = Ry (1- \frac{1}{n^2}),
\eeqa
where $Ry$ is the Rydberg constant and $z_0$ is the internal partition function of the hydrogen atom.  The electrons, ions and atoms have internal degrees of freedom, which affect their partition functions.

\begin{figure}
 \centering
\includegraphics[angle=0,width=5cm]{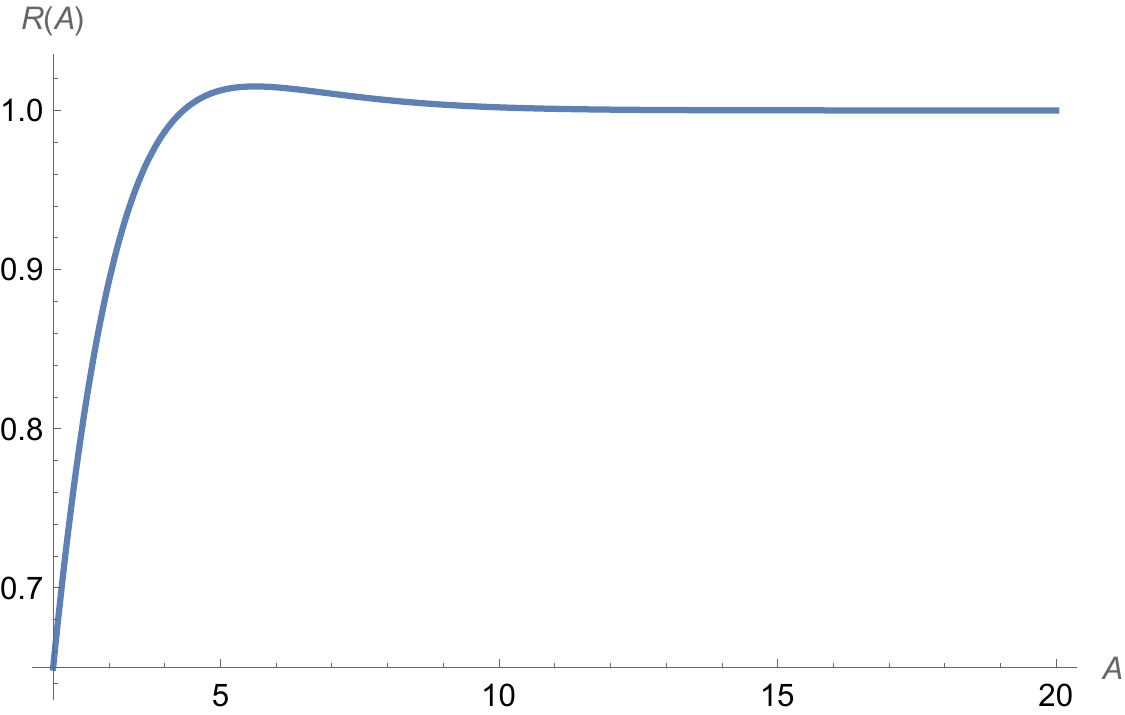} ~~
\centering
\caption{ This curve demonstrates the behaviour of the ratio $\CR(A)$ in (\ref{ratiotoren}) as a function of the  parameter $A= Ry/k T$.   To the value $A=1$ corresponds the temperature $T_1= 157961 K$, while to $A=r$ corresponds $T_r=T_1/r$.  At the low temperatures $ T< T_{20} = 7898 K$,  the ratio tends to one $\CR \approx 1$.  In the vicinity of $ T_{5}= 31592 K$  the $\CR $ is larger than one. The maximal deviation is about  3\%,  meaning that the ionisation is smaller than predicted by the Saha equation.  While at $T> T_3 = 52654 K$ the degree of ionisation is much larger than predicted by the Saha equation. }
\label{fig1} 
\end{figure}

 This partition function diverges because of the high degeneracy $z_0 n^2$ of the hydrogen atom energy states. We are suggesting  the following epsilon $\exp{(-\epsilon n)}$ regularisation of the $ZI_{bare}$:
\beqa\label{epsilonreg}
ZI(\epsilon) = \sum^{\infty}_{n=1} z_0 n^2 e^{  -\frac{Ry }{k T} (1- \frac{1}{n^2})  - \epsilon n} = 
e^{  -\frac{Ry }{k T}  }  z_0 \sum^{\infty}_{n=1}  n^2 e^{  \frac{Ry }{k T n^2}    - \epsilon n},
\eeqa
therefore we have to calculate the following sum:
\beqa 
  \sum^{\infty}_{n=1}  n^2 e^{  \frac{A}{n^2} - \epsilon n} &=&  \sum^{\infty}_{n=1}e^{ - \epsilon n} \sum^{\infty}_{m=0} \frac{A^m}{m!  ~n^{2m -2}} = \nn\\
 &=& \sum^{\infty}_{m=0} \frac{A^m}{m! }  \sum^{\infty}_{n=1}\frac{(e^{ - \epsilon})^n } {n^{2m-2}}= 
\sum^{\infty}_{m=0} \frac{A^m}{m! }    Li_{2m-2} (e^{ - \epsilon})  ,   
\eeqa
where we introduced a convenient parameter 
\be 
A= \frac{Ry }{k T} 
\ee
and the poly-logarithm function \cite{lewin1981, Sachdev:1993pr, Faddeev:1993rs} (see Appendix B)
\be
Li_s(x) = \sum^{\infty}_{n=1} \frac{x^n}{n^s }.
\ee
 The poly-logarithm function at $x=1$ reduces to the Riemann zeta function \cite{Savvidy:2018ffh}:
 \be\label{Reimann}
 \zeta(s)=\sum^{\infty}_{n=1} \frac{1}{n^s }.
 \ee  
 The poly-logarithm functions $Li_{-2}(x)$ and   $Li_{0}(x)$ are singular at $x=e^{ - \epsilon}=1$:
 \be
  Li_{-2}(x)=   \frac{ x(1+x)}{ (1-x)^3 } = \frac{2}{ \epsilon^3 } +\sum_{k=1}\CO(\epsilon^k), ~~~~~~~  Li_{0}(x)= \frac{ x }{(1-x)} = \frac{1}{ \epsilon }- \frac{1}{2} +\sum_{k=1}\CO(\epsilon^k),
\ee
and clearly identify the divergencies of the partition function:
\beqa\label{polylog}
&& \sum^{\infty}_{n=1}  n^2 e^{  \frac{A}{n^2} - \epsilon n} = \sum^{\infty}_{m=0} \frac{A^m}{m! }   Li_{2m-2} (e^{ - \epsilon})   \nn\\
&&= Li_{-2} (e^{ - \epsilon})  +  A ~Li_{0} (e^{ - \epsilon})  +  \sum^{\infty}_{m=2} \frac{A^m}{m! }   Li_{2m-2} (e^{ - \epsilon})\\
&&=\frac{2}{\epsilon^3} +A~ \Big(\frac{ 1}{\epsilon} - \frac{1}{2}\Big)  +  \sum^{\infty}_{m=2}     \zeta(2m-2) \frac{A^m}{m!} +\sum_{k=1}\CO(\epsilon^k)  \nn\\
 &&=\frac{2}{\epsilon^3} +A~ \Big(\frac{ 1}{\epsilon} - \frac{1}{2}\Big)  +\Big( \frac{\pi^2}{12} A^2 +\frac{\pi^4}{540} A^3  + \frac{\pi^6}{22680} A^4 +\frac{\pi^8}{1134000} A^5 +....+\sum^{\infty}_{k=1}\CO(\epsilon^k),~~~\nn
\eeqa
where $k=1,2,...$  This expression exposes  the structure of the divergencies, which are of the following finite type\footnote{The counterterms  are  of the zero and of the first order in $A$, and there is no need for  counterterms of higher order in $A$ representing a "renormalisable system" in terms of quantum field theory. }: 
\be\label{conterterms}
\frac{2}{\epsilon^3} + A~\Big(\frac{ 1}{\epsilon} -\frac{1}{2}\Big) + \sum^{\infty}_{k=1}\CO(\epsilon^k) =  \sum^{\infty}_{n=1}  n^2 e^{ - \epsilon n} +  A \sum^{\infty}_{n=1}  e^{ - \epsilon n}  .
\ee 
Therefore we can  renormalise the partition function by subtraction of the diverging terms and then taking the limit $\epsilon \rightarrow 0$. Thus the renormalised partition function takes the following form:
\beqa\label{neutral} 
 ZI&=&  z_0 \lim_{\epsilon \rightarrow 0}  e^{-A} \Big( \sum^{\infty}_{n=1}  n^2 e^{  \frac{A}{n^2} - \epsilon n} -  \frac{2}{\epsilon^3} - \frac{ A}{\epsilon}  + \frac{A}{2} \Big)   \nn\\
&=&z_0 e^{-A}  \Big( \frac{\pi^2}{12} A^2 +\frac{\pi^4}{540} A^3  + \frac{\pi^6}{22680} A^4 +\frac{\pi^8}{1134000} A^5 +...\Big) \nn\\
&=&  z_0 e^{-A}  \ \sum^{\infty}_{m=2}   ~ \zeta(2m-2) \frac{A^m}{m!} . 
\eeqa
In the low-temperature computation of the partition function it is approximated by the ground-state degeneracy  $ZI_{Saha} = z_0$, which is  assuming that most of the neutral hydrogen is in the ground state  (here, $z_0 =4$ is the statistical weight of the hydrogen ground state).  We will define the ratio 
\be\label{ratiotoren}
\CR = \frac{ZI}{ZI_{Saha}} = e^{-A}  \ \sum^{\infty}_{m=2}   ~ \zeta(2m-2) \frac{A^m}{m!} 
\ee
measuring the deviation of the renormalised partition function from the Saha low-temperature approximation. Its qualitative behaviour as a function of temperature is shown in Fig.\ref{fig1}.  In the temperature interval between 0 and 10000K the deviation is smaller than 
$0.003\%$, while between 10000 K and 34300 K the ratio increases from $0.1\%$ to $2.9\%$ and  drops  to $0.1\%$ at 46200K. This means that in the temperature interval between 10000K  to 46 200 K the renormalised partition function (\ref{neutral}) predicts a smaller  degree  of ionisation. From 46 200 K  until 100 million K and higher temperatures the Saha equation predicts much less degree of ionisation than the renormalised equation. Physically this is easy to understand because as the temperature increases, the fraction of neutral atoms is diminishing, while the Saha partition function of neutral atoms remains constant.  At these extreme temperatures the electrons are still non-relativistic as far as the temperature is less than $5.9 \times 10^{9} K$. The above results are applicable to the deuterium, tritium and hydrogen-type systems.

The renormalisation of the partition function induces the renormalisation of the bare probability weights $ n^2 e^{- A (1-\frac{1 }{n^2} )}$ in  (\ref{bare}).  Indeed, the right-hand side of the expression (\ref{conterterms}) allows to define a term-by-term renormalisation of the bare probability weights, so that 
the fraction of neutral hydrogen atoms on the energy level $n$ we will define in the following form:
\beqa\label{Gibbspn}
P_n(  \epsilon )&=&  
 \frac{  e^{  -A  }( n^2 e^{ \frac{A }{n^2} }  -n^2  -  A ) e^{  -  \epsilon n } }{e^{-A} \Big( \sum^{\infty}_{n=1}  n^2 e^{  \frac{A}{n^2} - \epsilon n} -  \frac{2}{\epsilon^3} - \frac{  A}{\epsilon}  +\frac{A}{2} \Big)}. 
\eeqa
Importantly, it is always a positive real function. One can check that these probabilities sum up to 1: 
\be\label{sumuptoone}
 \lim_{\epsilon \rightarrow 0}   \sum^{\infty}_{n=1}  P_n(\epsilon)  = 1.
 \ee
In the limit $ \epsilon \rightarrow 0$ the probability $P_n$ will take the following form:
\be\label{nstateprob}
P_n  =  \frac{ \sum^{\infty}_{m=2}   \frac{A^m}{m!}  \frac{1}{n^{2m-2}}}  { \sum^{\infty}_{m=2}    ~ \zeta(2m-2) \frac{A^m}{m!}}.
\ee
In calculating the partition functions of the atoms and ions in dissociative equilibria it is essential to measure all energies from the same level, for example, from the ground level of the neutral atom.  Therefore one should add 
 the ionisation energy $Ry$ to each level of the ion.  The partition function of ionised states will be identical with  the Saha equation \cite{Saha-1920,Saha-1920-a, Saha-1921,Saha_1921-b,  Payne1925, 10.1093/mnras/stz549, Blanco2014, Eisenstein_2011,Majewski_2017}:
\beqa\label{ionised}
ZII &=& z_i z_e \int e^{  -\frac{  Ry  + \frac{p^2}{2m} }{k T}     }  \frac{d^3 x d^3p}{(2 \pi \hbar)^3}  
=\frac{z_i z_e }{n_e}  \Big(\frac{  2 \pi m k T } {(2 \pi \hbar)^2 }  \Big) ^{3/2}~ e^{  - \frac{Ry}{ k T}  }, 
\eeqa
where $n_e = 1/V_e$ is the number of electrons in a unit volume and $z_i$, $z_e$ are internal  partition functions of ion and electron. 
 
The   partition function of $H, H_+, e^-$ plasma is the product of the partition functions of the individual contributions of electrons and  of the neutral and ionised hydrogen states:
\beqa\label{total}
Z &=&  ZI  \cdot ZII = z_0 e^{-A} \sum^{\infty}_{m=2}  ~ \zeta(2m-2) \frac{A^m}{m!}   ~ \cdot \frac{z_i z_e}{n_e}  \Big(\frac{  2 \pi m k T } {(2 \pi \hbar)^2 }  \Big) ^{3/2}~ e^{  - \frac{Ry}{ k T}  } .~~
\eeqa
Let us consider the low-temperature limit of the partition function when $T \rightarrow 0$ and $A \rightarrow \infty$.  In the limit of large real values of $s = 2m -2$ the Riemann zeta function converges to one:
\be
 \lim_{s \rightarrow \infty } \zeta(s)  \rightarrow 1, ~~~~~
\ee
therefore 
\beqa\label{limitlow}
&&\sum^{\infty}_{m=2}    \zeta(2m-2) \frac{A^m}{m!} ~~\approx ~~ \sum^{\infty}_{_{m=2}} \frac{A^m}{m!} ~~\approx ~~ e^{A},
\eeqa
and the low temperature partition function  $Z $  is  
\be\label{Sahalimit}
Z_{Zaha}   =   \frac{z_0 z_e  z_i }{n_e}  \Big(\frac{  2 \pi m k T } {(2 \pi \hbar)^2 }  \Big) ^{3/2}  e^{  -\frac{Ry  }{k T}  } .
\ee
This is a standard  expression of the Saha partition function and is used to calculate the radiance  of the spectral lines and of other physical properties of plasma \cite{Saha-1920,Saha-1920-a, Saha-1921,Saha_1921-b, Fowler1,Fowler2, Payne1925, 10.1093/mnras/stz549, Blanco2014, Eisenstein_2011,Majewski_2017}.  It was obtained by approximating the partition function of neutral atoms $ZI$ equal to the ground state degeneracy $ZI_{Saha} = z_0$, while $ZII$ has its canonical  form. 

The renormalisation of the  partition function and of the probability weights  allows to calculate the thermodynamical characteristics of the hydrogen gas in this new setting. The energy mean value of the neutral atoms and the variance are defined as the energy averages with respect to the renormalised probability weights (\ref{nstateprob}):
\beqa\label{meanvalues}
&&U= \langle EI \rangle   = \sum^{\infty}_{n=1}  Ry  \Big(1- \frac{1}{n^2} \Big)   \sum^{\infty}_{m=2} \frac{1}{m!} \Big(\frac{Ry}{k T}\Big)^m \frac{z_0}{n^{2m-2}}/ZI   \label{menEI}\\
&&\langle EI^2 \rangle   = \sum^{\infty}_{n=1}  Ry^2  \Big(1- \frac{1}{n^2} \Big)^2   \sum^{\infty}_{m=2} \frac{1}{m!} \Big(\frac{Ry}{k T}\Big)^m \frac{z_0}{n^{2m-2}}/ZI  \label{menEIsq}\\
&&(\triangle EI)^2 =  \langle EI^2 \rangle - \langle EI \rangle^2 .
\eeqa
The energy mean value  (\ref{menEI}) can be expressed as a partial derivative of the renormalised partition function $ZI$ but with the additional term that appears due to the  renormalisation of the  probability weights :
 \beqa\label{meanvalues1}  
\langle EI \rangle   = k T^2 \frac{1}{ZI} \frac{\partial ZI}{\partial T} + \frac{\pi^2}{3 ZI}    \frac{Ry^2}{k T}  e^{  -\frac{Ry  } {k T}} .
 \eeqa
Similarly, the mean value of the energy square can be expressed as a second derivative of the partition function.

\section{\it Fraction of  ionised atoms in hydrogen plasma}

In order to calculate the optical properties of an ionised hydrogen gas and of the other physical properties of the hydrogen plasma one should know the fraction of ionised atoms relative to the neutral ones  when the hydrogen plasma is in dissociative equilibria.  The fraction can be expressed as the following ratio   \cite{Saha-1920,Saha-1920-a, Saha-1921,Saha_1921-b, Fowler1,Fowler2}:
\beqa\label{equilib}
 \frac{n_{II}  }{n_{I}}&=& \frac{ZII}{  ZI},   
\eeqa
where $n_{I}$ and  $n_{II}$   are the number densities of neutral atoms and ions. By substituting the  expressions of the corresponding partition functions (\ref{neutral} ) and (\ref{ionised}) we will obtain  that the renormalised ionisataion equation
\beqa\label{fractionHII}
 \frac{n_{II} n_e }{n_{I}}&=&  \Big(\frac{  2 \pi m k T } {h^2 }  \Big) ^{3/2}  \Big( \frac{z_e  z_i}{z_0 }  \Big) \frac{1
}{   \sum^{\infty}_{m=2}   \zeta(2m-2) \frac{1}{m!} \Big( \frac{Ry}{k T} \Big)^m} ~.
\eeqa
In the ideal gas approximation the electron number density  $n_e$ can be expressed in terms of pressure and temperature, $P_e n_e = k T$, so that the ratio will take the following form:
\beqa\label{fractionHII1}
 \frac{n_{II}}{n_{I}}&=&\frac{  k T}{P_e}   \Big(\frac{  2 \pi m k T } {h^2 }  \Big) ^{3/2}  \Big( \frac{z_e  z_i}{z_0 }  \Big) \frac{1
}{   \sum^{\infty}_{m=2}    \zeta(2m-2) \frac{1}{m!} \Big( \frac{Ry}{k T} \Big)^m}~ .
\eeqa
This equation  shows that pressure has a  great influence on the degree of ionisation. A reduction in the pressure value is attended with greatly enhanced ionisation.  

Now one can compare renormalised ionisation equation (\ref{fractionHII}) with the Saha ionisation equation.  The Saha equation has the following form \cite{Saha-1920,Saha-1920-a, Saha-1921,Saha_1921-b, Fowler1,Fowler2}:
\beqa\label{sahafractionHII}
\Big( \frac{n_{II} n_e}{n_{I}}\Big)_{Saha}&=&\ \Big(\frac{  2 \pi m k T } {h^2 }  \Big) ^{3/2}  \Big( \frac{z_e  z_i}{z_0 } \Big) e^{  -\frac{Ry  }{k T}  },
\eeqa
and the deviation from the Saha equation is defined by the ratio  
\be\label{ratiotoren1}
\CR=  \Big( \frac{n_{II} n_e}{n_{I}}\Big)_{Saha}/  \Big(\frac{n_{II} n_e }{n_{I}} \Big) = e^{  -\frac{Ry  }{k T}  }   \sum^{\infty}_{m=2}   \zeta(2m-2)   \frac{ 1 }{m!}\Big( \frac{Ry}{k T} \Big)^m,
\ee
which we alredy introduced above (\ref{ratiotoren}). The figure Fig.\ref{fig1} demonstrates  the dependence of the ratio $\CR$  as a function of the  $A =  \frac{Ry  }{k T} $. The Saha equation \cite{Saha-1920,Saha-1920-a, Saha-1921,Saha_1921-b,Fowler1,Fowler2} appears here as a low-temperature approximation.  It follows that the exact result for the partition function (\ref{total}) describes the physics in a larger temperature interval  and the deviation from the standard expressions  at higher temperatures.

It is important to know  the fraction of  ionised atoms $n_{II}$  relative to the total number of atoms $n = n_{I}+n_{II}$.  
It can be defined in the following way: 
$$
\frac{n_{II}}{n_{I}} = \frac{n_{II}}{n-n_{II}} = \frac{x}{1-x}, 
$$
where 
$$
x = \frac{n_{II}}{n}
$$ is the fraction of ionised atoms\footnote{
It is also true that $ x = \frac{ZII}{  ZI +ZII} $, while the fraction of neutral atoms is $1-x = \frac{ ZI}{ ZI +ZII} $.}. From the equation (\ref{fractionHII})  we have
\beqa
n_e   =   \frac{1-x}{x}     \Big(\frac{  2 \pi m k T } {h^2 }  \Big) ^{3/2}  \Big( \frac{z_e  z_i}{z_0 }  \Big) \frac{1
}{   \sum^{\infty}_{m=2}   \zeta(2m-2) \frac{1}{m!} \Big( \frac{Ry}{k T} \Big)^m}.
\eeqa
In the ideal gas approximation, $P_e n_e = k T$,  the equation will take the following form
\beqa\label{pressure}
P_e = \frac{1-x}{x}   k T   \Big(\frac{  2 \pi m k T } {h^2 }  \Big) ^{3/2}  \Big( \frac{z_e  z_i}{z_0 }  \Big) \frac{1}{   \sum^{\infty}_{m=2}   \zeta(2m-2) \frac{1}{m!} \Big( \frac{Ry}{k T} \Big)^m}.
\eeqa
Expressing the partial number densities through the total number denisity $n$
$$ 
\frac{n_{II} n_e }{n_{I}}= n \frac{x^2}{1-x}
$$
the degree  of ionisation $x$ can be found by using the renormalised equation  (\ref{fractionHII}) 
\beqa
\frac{x^2}{1-x} = \frac{1}{n}~\Big( \frac{z_e  z_i}{z_0 }  \Big)     \Big(\frac{  2 \pi m k T } {h^2 }  \Big) ^{3/2} \frac{1}{   \sum^{\infty}_{m=2}   \zeta(2m-2) \frac{1}{m!} \Big( \frac{Ry}{k T} \Big)^m}.
\eeqa
The low-temperature limit (\ref{limitlow}) of this equation reduces to the Saha equation
\beqa\label{Sahax}
\frac{x^2_S}{1-x_S} = \frac{1}{n}~\Big( \frac{z_e  z_i}{z_0 }  \Big)     \Big(\frac{  2 \pi m k T } {h^2 }  \Big) ^{3/2}  e^{  -\frac{Ry  }{k T}  }.
\eeqa
As an example let us consider the {\it solar core} with  $n= 10^{26} cm^{-3}$, $T = 10^{7} K$, then 
\be
x \approx 0.9997
\ee
and the hydrogen plasma is fully ionised, while the Saha equation predicts a partial ionisation 
$$
x_S \approx 0.56.
$$
In {\it solar photosphere} with $n = 10^{16} cm^{-3}$, $T = 6000 K$ we will get 
$$x= 6.4 \times 10^{-4}$$ and this degree of ionisation  coincides with the value predicted by the Saha equation.  The last coincidence of the degree of ionisation is not accidental, it follows from the $\CR(T)$ behaviour, the low-temperature behaviour of both equations is identical. 

The fraction of neutral atoms  in the r-state $ \frac{N_r}{NI}$ suitable for absorption of  light  can also be computed and is equal to the population of the corresponding energy level $P_r$ (\ref{nstateprob}):
 \be\label{nstateprob1}
\frac{N_r}{NI}=P_r  =  \frac{   \sum^{\infty}_{m=2}  \frac{1}{r^{2m-2}}  \frac{1}{m!} \Big( \frac{Ry}{k T} \Big)^m   }  { \sum^{\infty}_{m=2}    ~ \zeta(2m-2) \frac{1}{m!} \Big( \frac{Ry}{k T} \Big)^m  }.
\ee
The fraction  $n_r = \frac{N_r}{N} $ of the total number of atoms that are neutral and in the r-state is then 
 \be\label{ournr}
 n_r = P_r (1-x).
 \ee
 From equation (\ref{pressure}) we have 
 \be
 1-x = \frac{ZI}{ ZI + ZII}=\frac{\sum^{\infty}_{m=2}    ~ \zeta(2m-2) \frac{1}{m!} \Big( \frac{Ry}{k T} \Big)^m }{\sum^{\infty}_{m=2}    ~ \zeta(2m-2) \frac{1}{m!} \Big( \frac{Ry}{k T} \Big)^m  + \frac{1}{n_e} \Big( \frac{z_e  z_i}{z_0 }  \Big) \Big(\frac{  2 \pi m k T } {h^2 }  \Big) ^{3/2} }. 
 \ee
Hence from (\ref{ournr}) it follows that 
\beqa\label{populationr}
n_r= \frac{ \sum^{\infty}_{m=2}  \frac{1}{r^{2m-2}}  \frac{1}{m!} \Big( \frac{Ry}{k T} \Big)^m  }{\sum^{\infty}_{m=2}   ~ \zeta(2m-2) \frac{1}{m!} \Big( \frac{Ry}{k T} \Big)^m  +  \frac{1}{n_e}   \Big( \frac{z_e  z_i}{z_0 }  \Big)\Big(\frac{  2 \pi m k T } {h^2 }  \Big) ^{3/2}  }.
\eeqa
As an example the population of the first four energy levels of neutral hydrogen atoms at the photosphere temperature $T=6000 K$ will be 
\be
n_1  \approx 0.999355,~~   n_2   \approx 1.062 \times 10^{-8}, ~~n_3 \approx 5.8445 \times 10^{-10}, ~~n_4 \approx 2.4643 \times 10^{-10}. 
\ee
In the case of Saha equation the population of energy levels is defined by the equation
\be
n_{Sr} = \frac{ e^{ \frac{Ry}{k T} \frac{1}{ r^2} } }{ e^{\frac{Ry}{k T }}  + \frac{1}{n_e}  \Big( \frac{z_e  z_i}{z_0 }  \Big)\Big(\frac{  2 \pi m k T } {h^2 }  \Big) ^{3/2} },
\ee
where the density of electrons $n_e= n x_S$ can be expressed through the total number density $n$ using (\ref{Sahax}). The population of the first four energy levels of neutral hydrogen atoms at the photosphere temperature $T=6000 K$ will be 
 \be
n_1  \approx 0.999355,~~   n_2   \approx 1.063 \times 10^{-8}, ~~n_3 \approx 6.17599 \times 10^{-10}, ~~n_4 \approx 3.05338 \times 10^{-10}
\ee
and for the low lying energy levels the difference with renormalised equation is very small.  The difference 
increases for the higher energy levels 
\beqa
&&n_{100} \approx   9.70559 \times 10^{-11},~~     n_{1000} \approx 9.69295 \times 10^{-11},~~  n_{100000} \approx 9.69282 \times 10^{-11}\nn\\
&&n_{S100} \approx   3.69198 \times 10^{-8},~~     n_{S1000} \approx 3.68187 \times 10^{-6},~~     n_{S100000} \approx 0.0369177.
\eeqa
The differences that appeared at the higher energy levels with the Saha equation are understandable and are the result of the divergency of the partition function that does not allow to normalise the sum of the probabilities to one. The renormalised partition function defines the probabilities of energy states that sum up to one (\ref{sumuptoone}), (\ref{nstateprob1}). 

The fractions of excited atoms $N_r$ on the energy levels $r$ to the atoms on the energy levels $n$ has the following form (\ref{nstateprob}):
\beqa\label{n2tototal}
&\frac{N_r}{N_n} = \frac{P_r}{P_n}  =   \frac{ \sum^{\infty}_{m=2}   \frac{A^m}{m!}  \frac{1}{r^{2m-2}}}{ \sum^{\infty}_{m=2}   \frac{A^m}{m!}  \frac{1}{n^{2m-2}}} .
\eeqa
The fraction of excited hydrogen atoms define the radiance of the absorption lines, and in particular  those of the Lyman, Balmer and Paschen series \cite{10.1093/mnras/stz549, Blanco2014, Eisenstein_2011,Majewski_2017,  Barrett1964SpectrumAR}.   The  absorption rate     $(r \rightarrow n)$ has the following form:
\be\label{absorp}
 \frac{(Absorption~rate~of~ r \rightarrow n)_{\gamma}} {N} = n_r~ \CD_{\gamma}(r \rightarrow n)  ~U_{\gamma} (\omega_{rn}),
 \ee
and depends on the number of atoms in the r-th state   $n_r$ given in (\ref{populationr}) and  of the photon absorption rate \cite{Dirac:1927dy,Fermi:1932xva, Fermi1954} (see Appendix B):
\be\label{photonabsorf1}
 \CD_{\gamma} (r \rightarrow n) = \frac{4 \pi^2 e^2}{3 \hbar^2} \vert \langle n \vert  x_i \vert r  \rangle \vert^2, 
\ee
while  $U_{\gamma} (\omega_{rn})$ is the spectral energy density (energy per unit volume and per unit frequency).   The  above formulas define the radiance of the absorption lines at which light of the background  source is diminished, producing dark lines on a relatively bright background  \cite{KRAMIDA2010586}.

\section{\it Conclusion }  
 
We suggested the epsilon regularisation (\ref{epsilonreg}) of the neutral hydrogen partition function and the corresponding  renormalisation (\ref{neutral} ) by adopting the methods that were well developed  in the quantum field theory and strings. The renormalised partition function $ZI(T)$ (\ref{mainpart}) is well defined at all temperatures and has a smooth low-temperature limit (\ref{limitlow}) and (\ref{Sahalimit}).  The  renormalisation of the  partition function allows to derive the renormalised  ionisation equation (\ref{fractionHII}) and (\ref{fractionHII1}), which is now well defined in a much larger temperature region and coincides with the standard Saha equation in the low-temperature limit, meaning below the temperatures of 10000 K.  

In order to measure the deviation of the renormalised ionisation equation (\ref{fractionHII}) from the classical Saha equation we  define the ratio (\ref{ratiotoren}), (\ref{ratiotoren1})
\be
\CR(T) =   \Big( \frac{n_{II} n_e}{n_{I}}\Big)_{Saha}/\frac{n_{II} n_e }{n_{I}}=e^{-A}  \ \sum^{\infty}_{m=2}   ~ \zeta(2m-2) \frac{A^m}{m!}. 
\ee
The behaviour of $\CR(T)$ as a function of temperature is shown in Fig.\ref{fig1}.  In the  temperature interval between 10000K  to 46 200 K the Saha equation (\ref{sahafractionHII}) predict a larger degree of ionisation than the renormalised equation (\ref{fractionHII}). From 46 200 K  until 100 million K the Saha equation predicts much less degree of ionisation than the renormalised equation. Physically this is easy to understand because as the temperature increases, the fraction of neutral atoms is diminishing, while the Saha partition function for neutral atoms remains constant.  At these extreme temperatures the electrons are still non-relativistic as far as the temperature is less than $5.9 \times 10^{9} K$. The above results are applicable to the deuterium, tritium and hydrogen-type systems.

The Saha equation has a wide range of applications in Astrophysics, Plasma  Physics, 
Spectroscopy, Material  Science, Cosmology, Astrochemistry and  the renormalised ionisation equation (\ref{fractionHII}) may become useful for the description of extreme temperatures that were reached in modern Tokamak fusion reactors working at self-sustaining temperature regime,  between 50 and 100 million K \cite{Tokamak_RevModPhys.70.537,TokamakChina}.  

\section{\it Acknowledgement  }
 
 GS would like to thank the School of Physics and Astronomy of the Sun Yat-sen
University for hospitality and support. GS would like also to thank Prof. Gevorg Hajyan for stimulation discussion of Saha ionisation equation. PZ was supported by the
National Natural Science Foundation of China (grant no. 12375084).

\begin{appendices}

\section{\it Divergency of hydrogen partition function}
 
 Hydrogen is the most abundant element in the Universe and its thermodynamic properties dominate the physical conditions in stellar atmospheres and in the interstellar/intergalactic medium and therefore requires  the description of its thermodynamical properties to be as accurate as possible in a wide range of temperatures and pressures intervals \cite{Saha-1920,Saha-1920-a, Saha-1921,Saha_1921-b, fowler1923dissociation, Fowler1,Fowler2, Payne1925,  Eisenstein_2011,10.1093/mnras/stz549, Blanco2014, Majewski_2017,1975BAAS....7..366E, 1957ApJ...125..742E,1965ZNatA..20..184D,1965dplt}.     The cosmic reionisation of hydrogen and helium  are also of a great interest in astrophysics and cosmology \cite{ Bunker2023,  2025A&A...693A..60H,  Kashino_2023,heintz2023,  2024ApJ...962...24S, Zier:2026mem}.  
The description of the thermodynamical properties of hydrogen gas was facing immense uncertainty that is caused  by the divergency of the partition function arising due to the fast growing degeneracy factor $n^2$ of its excited n-state  \cite{Planck,Fowler1,Fowler2, Urey1924, Fermi1, Fermi2, 1986ApJ...310..723R,1936Phy.....3..729U, PhysRev.94.77, Larkin, Rouse_1983A, 1982ApJ...258..404U,PLASTINO2019122169,KRAMIDA2010586}.

Two  approaches to handle  the divergency problem were considered in literature and are based on the analysis of the interaction between hydrogen atoms and between atoms and electrons. The interaction between atoms appears when the size of the high excited states became comparable  with mean distance between atoms.  Indeed, the mean value of the electron radii  increases as $n^2  a $ in the excited n-state, where $a$ is the Bohr radius.  For large $n$ this radius becomes comparable to the average nearest-neighbour distance between atoms,  $n^2 a \approx n^{-1/3}$, where $n$  is the number density of atoms in unit volume, and it points to the fact that one should take into account the interaction of atoms. This interaction can limit the number of observable hydrogen lines  in a stellar atmosphere and may define the energy at which the summation   in calculating the partition function should be cut off.  The effective Urey-Fermi cut off function \cite{Urey1924,Fermi1} is 
 $$
 c_n(T)=e^{-b_n \frac{P}{R T}},
 $$ 
 where $b_n \approx n^6 a^3$ is 4 times the volume occupied by a mol of atoms in an excited n-state,  the quantity that is naturally entering  the van der Waals equation of the state $P(V-b_n) = R T$\footnote{The deviations from the law of the perfect gas is   due to the interaction of atoms that changes the volumes occupied  by  the atoms  \cite{Urey1924,Fermi1}.}.  The effective partition function would converge due to the $n^6$ factor in the exponent\footnote{It is the first example of the effective partition function calculation in the statistical mechanics by Fermi \cite{Fermi1} and is a prototype of a similar calculation of the effective actions in Quantum  Electrodynamic (QED) and Quantum Chomodynamics (QCD) \cite{Heisenberg:1936nmg,Schwinger:1951nm,Savvidy:2019grj}. }:
 \be
 Z_{eff} \approx \sum^{\infty}_{n=1} z_0 n^2 e^{-\frac{Ry}{kT}(1-\frac{1}{n^2}) -n^6a^3 \frac{P}{R T} }.
 \ee
In an alternative approach  the problem  was considered through the interaction of scattered  electrons with the hydrogen atoms, leading to the Planck-Larkin cutoff function \cite{Planck,Larkin}.  This interaction induces  the deformation of the Coulomb potential in a partially ionised gas where the electrons and ions interact with one another.  The electric potential in the space between charged particles becomes slightly lowered beneath the theoretical zero potential for an isolated charge, and {\it this eliminates the infinite number of Rydberg levels just beneath the ionisation limit, so that there remains only a finite number of levels and a finite number of terms in the partition function} \cite{Planck, Larkin, Rouse_1983A, 1987ApJ...319..195D, 1982ApJ...258..404U, PLASTINO2019122169, KRAMIDA2010586}.  The corresponding Planck-Larkin cutoff function that appears in that case has elements that are similar in their structure to the one obtained in this article, but has different physical origin.  In our approach the partition function is considered  for a  single atom, its spectrum remains untouched and contains an infinite number of bound states,  while  the interaction is considered by the calculation of the partition function for the whole system.

\section{\it   Poly-logarithm function}

For $\vert \epsilon \vert  \leq 2 \pi$  the poly-logarithm function $ Li_{2n}(e^{-\epsilon})$ has the following representation \cite{110}:
\beqa\label{series1}
 Li_{2n}(e^{-\epsilon})= \sum^{\infty}_{k=0,~k \neq 2n-1} \frac{(-\epsilon)^k}{k!} \zeta(2n-k)
 - \frac{\epsilon^{2n-1}}{(2n-1)!}[H_{2n-1} - \ln{\epsilon}],
 \eeqa
where $n=1,2,...$ and 
 \be
 H_n= \sum^{n}_{m=1}  \frac{1}{m}, ~~~~~~~~H_0 =0.
 \ee
All terms have positive power of $\epsilon$ except for the  $k=0$  term in the sum, therefore the series expansion in $\epsilon$ has the following form:
\beqa\label{series2}
 Li_{2n}(e^{-\epsilon})=  \zeta(2n)  + \frac{\epsilon^{2n-1}}{(2n-1)!}   \ln \epsilon +  \sum^{\infty}_{k=1} c_k \epsilon^k 
\eeqa  
 and in the limit $ \epsilon \rightarrow 0$
\beqa\label{series3}
\lim_{\epsilon \rightarrow 0 } Li_{2n}(e^{-\epsilon})=  Li_{2n}(1) = \zeta(2n), ~~~~n=1,2,...   
\eeqa

 \section{\it Photon  absorption rates   }

The absorption rate of the photons is defined by the expression (\ref{photonabsorf1}) and for the low-level energy states has the following form:
\beqa\label{photonabsor1}
& \CD_{\gamma}(1s \rightarrow 2p) = \frac{ 2^{17}}{3^{10}} \Big(\frac{e^2}{\hbar c}\Big)^{-1}   \Big(\frac{ \pi^2 \hbar }{ m^2 c}\Big) ,~~~~   \CD_{\gamma}(1s \rightarrow 3p) = \frac{3^6}{2^{11}}  \Big(\frac{e^2}{\hbar c}\Big)^{-1}   \Big(\frac{ \pi^2 \hbar }{ m^2 c}\Big), \nn\\
& \CD_{\gamma}(2p  \rightarrow 3d) = \frac{3^6 2^{25}}{5^{13}}  \Big(\frac{e^2}{\hbar c}\Big)^{-1}   \Big(\frac{ \pi^2 \hbar }{ m^2 c}\Big),~~~~ 
 \CD_{\gamma}(2s \rightarrow 3p ) = \frac{3^6 2^{22}}{5^{12}}  \Big(\frac{e^2}{\hbar c}\Big)^{-1}   \Big(\frac{ \pi^2 \hbar }{ m^2 c}\Big) ,~\nn\\
& \CD_{\gamma}(2p  \rightarrow 3s) = \frac{3^6 2^{17}}{5^{12}}  \Big(\frac{e^2}{\hbar c}\Big)^{-1}   \Big(\frac{ \pi^2 \hbar }{ m^2 c}\Big), ......
 \eeqa
with the following numerical values:
 \beqa\label{photonabsor2}
& \CD_{\gamma}(1s \rightarrow 2p)  \approx 1.27 \times 10^{20} \frac{cm}{gram},~~~~~~ \CD_{\gamma}(1s \rightarrow 3p)  \approx 2.04 \times 10^{19} \frac{cm}{gram}, \nn\\
& \CD_{\gamma}(2p \rightarrow 3d)  \approx 1.15 \times 10^{21} \frac{cm}{gram},~~~~~~  \CD_{\gamma}(2s \rightarrow 3p)  \approx 7.18 \times 10^{20} \frac{cm}{gram},  \nn\\
&\CD_{\gamma}(2p \rightarrow 3s)  \approx 2.24 \times 10^{19} \frac{cm}{gram},....
 \eeqa

The  fractional absorption rate  $(n \rightarrow m)$ has the following form:
\be\label{absorp}
 \frac{(Absorption~rate~of~ r \rightarrow n)_{\gamma}} {N} = n_r \CD_{\gamma}(r \rightarrow n)  ~U_{\gamma} (\omega_{rn}),
 \ee
 where the ratio $n_r $ is given in (\ref{populationr}).




\end{appendices}

\bibliography{gravityatom}

\end{document}